# The Smallest Metallic Nanorods Using Physical Vapor Deposition

Xiaobin Niu[1,3], Stephen P. Stagon[1,3], Hanchen Huang[1,*], J. Kevin Baldwin[2] and Amit Misra[2]

[1] Department of Mechanical Engineering, University of Connecticut, Storrs, CT 06269

[2] MS K771, Los Alamos National Laboratory, Los Alamos, NM 87545

[3] The two authors contributed equally to this work

[*] E-mail: hanchen@uconn.edu

**The growth of metallic nanorods using physical vapor deposition (PVD) allows for great control of chemical composition, crystalline structure, and even architecture[1,2]. As the linear dimension shrinks to ~10 nm, metallic nanomaterials start to have unique properties, such as catalytic function or low-temperature melting, which their bulk counterparts do not have[3-6]. However, PVD growth of small metallic nanorods – with a diameter of ~10 nm – has not been reported. Due to the lack of a theoretical foundation of nanorod growth, the physical limit of the smallest diameter is unknown. As a result, the pursuit of the smallest nanorods has no clear target, and consequently no clear path to the target. Here, we first present a theory of the smallest diameter and then use the theory to guide the experimental realization of Cu nanorods of ~20 nm in diameter and Au nanorods of ~10 nm in diameter,** *the smallest well-separated metallic nanorods ever reported using PVD.*

In order to push the limit of the smallest nanorods, experimentation alone is insufficient and it becomes effective when coupled with a theoretical understanding of the physical principles that set the limit. In this Letter, we present (1) a closed-form theory of the smallest diameter, (2) verification of the theory using lattice kinetic Monte Carlo (LKMC) simulations and validation using previous experiments, and (3) realization of the smallest nanorods using theory-guided PVD experiments.

For the theoretical formulation, the conceptual framework of nanorod growth serves as the starting point[7]. In contrast to the theories for the growth of large crystals instead of crystalline nanorods[8,9], this framework recognizes that multiple-layer surface steps are kinetically stable[10]; in contrast, the classical theory predicts that such steps are kinetically unstable[11]. Further, these multiple-layer surface steps dictate the diffusion of adatoms during nanorod growth[12-14]. Under this framework, metallic nanorods grow in two modes – I and II (Fig. 1). In mode I, the growth takes place on wetting substrates and nanorods have the shape of a tower[15]. The competition between multiple-layer and monolayer surface steps defines the diameter of nanorods, and also defines the slope on the side of nanorods. The diameter becomes smaller if more of the surface steps are multiple-layer instead of monolayer. In mode II, the growth takes place on non-wetting substrates and nanorods have the shape of a cylinder (or of an inverted tower if they grow sufficiently tall). Because of the complete, or nearly complete, dominance of multiple-layer surface steps over monolayer surface steps, growth model II results in the smallest diameter of nanorods.

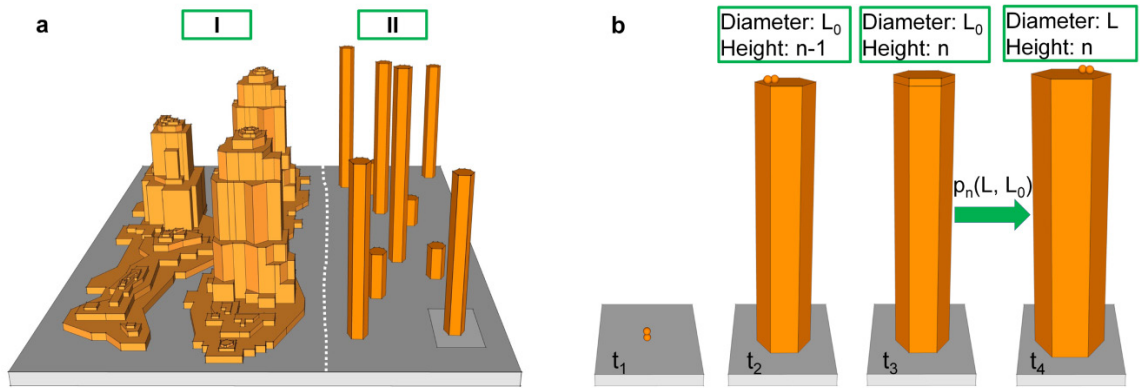

**Figure 1: Model of growing the smallest nanorods. a**, Schematic of the two modes of nanorod growth, with mode II giving rise to the smallest nanorods; and **b**, evolution of a nanorod as a function of time for mode II.

Focusing on growth mode II, we first describe our physical model of nanorod growth; the mathematical formulation then turns the model into a closed form theory.

The model starts with nucleation on a non-wetting substrate (snapshot $t_1$ in Fig. 1b). Due to non-wettability, the critical size of nucleating the second layer is one atomic diameter. As the nanorods grow, they have the shape of cylinder (snapshot $t_2$ in Fig. 1b). Since the diameter of the nanorods is small, only one adatom will be on top most of the time, and a new layer nucleates once two adatoms present simultaneously; this is also called the lone adatom model (LAM)[9]. The snapshot $t_2$ in Fig. 1b shows the configuration with the nucleus of a new layer. Aiming at the smallest diameter, we consider the complete geometrical shadowing condition – that is, atoms are deposited onto only the top of nanorods, not onto the sides. With the small diameter of nanorods and the large diffusion barrier at the multiple-layer steps or edges of the nanorods, the newly nucleated layer will grow to full coverage before any deposited atoms diffuse to the side. The snapshot $t_3$ in Fig. 1b shows the configuration when the coverage of one layer completes. The snapshot $t_4$ in Fig. 1b is similar to the snapshot $t_2$, except with one extra layer on top of the nanorod.

Based on the physical model of nanorod growth, the clock in our theoretical formulation starts at the moment when the coverage of the $n^{th}$ layer has just been completed (snapshot $t_3$ in Fig. 1b). The cross-sectional area is $A = \alpha L^2$ with $L$ being $L_0$ at this moment. The $\alpha$ is a geometrical factor; $\alpha = \pi/4$ for circular cross-sections and $\alpha = 1$ for square across-sections. For easy comparison with experiments, we will refer to $L$ as the "diameter", even though it is precisely diameter only for circular cross-sections. Before the next layer is nucleated, mass conservation requires $\int_0^t F_e \alpha L^2 dt = n\alpha L^2 - n\alpha L_0^2$; $F_e$ is the effective deposition rate on top of the nanorod, and $t$ is the time. It should be noted that this conservation equation is valid for mode II of nanorod growth in Fig. 1a, and that it is different from that for the growth of large crystals[8,9].

Using the conservation equation and following the LAM, we derive the distribution $f_n(L, L_0) = 1 - \exp\left[(L_0^5 - L^5)/L_n^5\right]$ as the fraction of nanorods on top of which nucleation has taken place when the diameter of nanorods is $L$. Here, $L_n = \left[(10v_{3D})/(n\alpha^2 F_e)\right]^{1/5}$ and $v_{3D}$ is the diffusion jump rate of adatoms over multiple-layer surface steps. The nucleation probability density that the $(n+1)^{th}$ layer starts to nucleate on top of a nanorod of diameter $L$ is then $p_n(L, L_0) = df_n(L, L_0)/dL = \left\{5L^4 \exp\left[(L_0^5 - L^5)/L_n^5\right]\right\}/L_n^5$.

Next, we consider the fact that not all nanorods have the same diameter $L_0$ at snapshot $t_2$ in Fig. 1b. Instead, if their size distribution is $S_{n-1}(L)$, the size distribution at snapshot $t_4$ is $S_n(L) = \int_0^L dl S_{n-1}(l) p_n(L, l)$. For a non-wetting substrate, we approximate the size distribution of the first layer as a delta function, $S_1(L) = \delta(L - 0)$. With this approximation, we recursively determine $S_n(L)$. Finally, we determine the peak diameter $L_{\min}$ as the $L$ that satisfies $dS_n(L)/dL = 0$. For a sufficiently narrow size distribution, this peak diameter $L_{\min}$ represents the smallest diameter. When the number of layers $n$ is large, we obtain a closed-form expression $L_{\min} \approx \left[(10/\alpha^2)\ln(n/2)(v_{3D}/F_e)\right]^{\frac{1}{5}}$. Since the effective deposition rate $F_e$ is proportional to the nominal deposition rate $F$ through $F_e = F \cdot \sin\theta$ with $\theta$ being the incidence angle, $L_{min} \propto (v_{3D}/F)^{1/5}$. (For detailed derivation, see Supplementary Information).

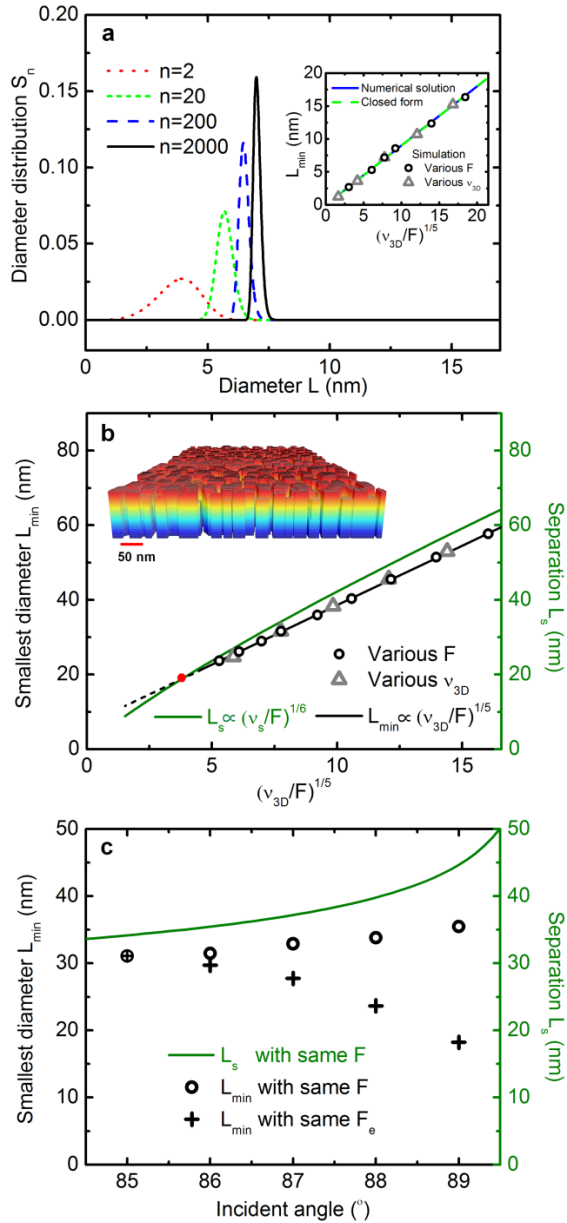

**Figure 2: Theory of the smallest diameter $L_{min}$ of nanorods. a**, The theoretical distribution $S_n(L)$ for various numbers of layers $n$ in height; the inset shows a comparison of the numerical solution, the closed-form expression, and LKMC simulation results under complete geometrical shadowing as a function of $(v_{3D}/F)^{1/5}$. **b**, LKMC simulation results under incomplete geometry shadowing as a function of $(v_{3D}/F)^{1/5}$; the separation of nanorod nuclei $L_s$ is included for comparison, and the incidence angle is 85°. The inset shows nanorods from a LKMC simulation with random nucleation. **c**, LKMC simulation results under incomplete geometry shadowing as a function of incidence angle, with either the same $F_e = 1 \cdot \sin 5°$ nm/s or the same $F = 1$ nm/s; the separation of nanorod nuclei $L_s$ is included for comparison.

Before using the theory, we verify it here. First, we numerically determine $S_n(L)$ as a function of the number of layers $n$ (effectively time). As Fig. 2a shows, the peak diameter first increases fast then more slowly with time, and the distribution become very narrow as $n$ reaches 2000 layers. The narrow distribution confirms the validity of using the peak diameter as representative of the smallest diameter $L_{min}$. Further, the numerical solution and the closed-form expression of $L_{min}$ are nearly identical as nanorods grow to 2000 layers (Fig. 2a inset). LKMC simulations using various substrate temperatures or various deposition rates, while keeping other conditions unchanged, show nearly identical dependence of $L_{min}$ on $(v_{3D}/F)^{1/5}$ as the theory predicts (Fig. 2a inset).

Upon verification of the theoretical formulations, we next use LKMC simulations to test the validity of the theory beyond complete geometrical shadowing conditions. As long as mode II of nanorod growth is operational, we still expect the dominance of multiple-layer surface steps, even if geometrical shadowing is incomplete. Indeed, the simulation results (Fig. 2b inset) show the dominance of multiple-layer surface steps. By changing $v_{3D}$ and $F$ independently, the simulation results show in Fig. 2b that $L_{min}$ is still proportional to $(v_{3D}/F)^{1/5}$ when the incidence angle of atomic flux is 85°. (For detailed simulation and verification setup, please see Supplementary Information).

Having verified the theory $L_{min} \approx \left[(10/\alpha^2)\ln(n/2)(v_{3D}/F_e)\right]^{\frac{1}{5}}$ and extended its applicability as $L_{min} \propto (v_{3D}/F)^{1/5}$ under incomplete geometrical shadowing, we now use a previous experiment[16] to validate it. In the experiment, Cu nanorods of ~30 nm in diameter grow under a deposition rate of 1 nm/s with an incidence angle of 85°; the substrate temperature is uncontrolled but is within 300-350 K. By increasing the deposition rate to 6 nm/s, the growth of nanorods transitions into the growth of a dense film. By including the theoretical separation of nanorod nuclei $L_s$ [17] in Fig. 2b, our theory explains this anomalous transition as the following. The crossover of $L_{min}$ and $L_s$ occurs at ~20 nm. As deposition rate increases, both $L_{min}$ and $L_s$ decrease. When they reach ~20 nm, $L_s$ becomes smaller than $L_{min}$, so there is no space for separate nanorods to exist. Because of random nucleation, some nanorods are separated at a smaller distance than the theoretical value $L_s$. As a result, nanorods bridge and merge even if $L_s > L_{min}$, provided they both are still close to ~20 nm. That is, $L_s$ makes it nearly impossible to grow well separated Cu nanorods that are smaller than ~30 nm; beyond our own experiments, others have also reported only nanorods of ~30 nm or larger but not smaller[18,19]. The fact that the theory explains the anomalous experimental results serves as a validation.

Now that the theory has been verified and validated, we use it to guide the pursuit of the smallest nanorods. The first insight from the theory is that $L_s$ is the limiting factor of growing smaller nanorods. If we can eliminate the constraint of $L_s$, it may become possible to grow smaller and well separated nanorods of diameter $L_{min}$. Putting this insight into action, we apply four strategies. (1) By using large incidence angles, we lower the effective deposition rate to promote the relationship $L_s > L_{min}$; (2) by using lower substrate temperatures, we take the advantage of larger activation energy in $L_{min}$ to promote the relationship $L_s > L_{min}$; (3) by using substrates with heterogeneous nucleation, we make $L_s$ is ineffective; and (4) by using highly non-wetting substrates, we increase $L_s$ to promote $L_s > L_{min}$. Since the last three strategies are apparent, we use Fig. 2c to show the feasibility of only the first strategy. As the incidence angle becomes larger, while keeping the nominal deposition rate constant, $L_{min}$ becomes larger but $L_s$ becomes even larger. Indeed, the increase of incidence angle promotes $L_s > L_{min}$.

The second insight is that a decrease of $v_{3D}$ (by an increase of the diffusion barrier of adatoms over multiple-layer surface steps) can be effective to reduce the diameter of nanorods according to $L_{min} \propto (v_{3D}/F)^{1/5}$. Putting this insight into action, we use quantum mechanics calculations to identify a metal with a large diffusion barrier of adatoms and therefore small $v_{3D}$. Our calculations show that the relevant energy barrier of adatoms diffusion down a multiple-layer surface step in Au is 0.52 eV, much larger than the 0.40 eV in Cu or 0.12 eV in Al[13,14]; this barrier is in contrast to the Ehrlich-Schwoebel barrier of adatoms diffusion down a monolayer surface step[11,20]. With this set of data, the second insight suggests that we can reach an even smaller diameter for Au nanorods than for Cu nanorods.

Using the first insight from the theory, we design the growth of Cu nanorods as the following. We use a large incidence angle of 88°, a substrate with heterogeneous nucleation sites of $SiO_2$, and a low substrate temperature of about 250 K. The experiments indeed confirm that well-separated Cu nanorods of ~20 nm in diameter grow (Fig. 3a), as the first theoretical insight suggests. This represents the smallest well-separated Cu nanorods that have ever been reported using PVD. Using both the first and the second insights from the theory, we grow Au nanorods using a large incidence angle of 88°, a substrate that is highly non-wetting (3M Copper Conductive

Tape 1182, 3M Corporation, St. Paul, MN), and a low substrate temperature of about 250 K. The experiments indeed confirm that well-separated Au nanorods of ~10 nm in diameter grow (Fig. 3b), as the two theoretical insights suggest. In fact, some of the Au nanorods are as small as 7 nm in diameter. Once again, the Au nanorods of ~10 nm in diameter are the smallest well-separated metallic nanorods that have ever been reported using PVD.

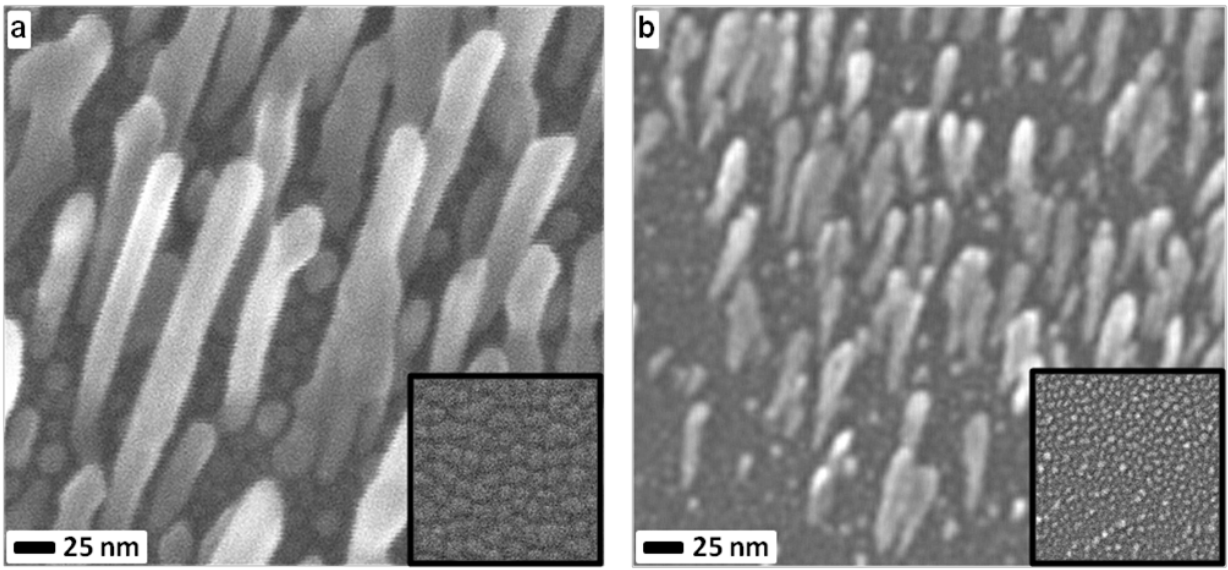

**Figure 3: Experimental results of the smallest well-separated nanorods.** Scanning electron microscopy (SEM) images of well-separated **a**, Cu and **b**, Au nanorods at an early stage; the insets with the same scale show the morphologies of substrates. The incidence angle of the deposition flux is 88°, the deposition rate is 0.1 nm/s, and the substrate temperature is estimated to be 250 K.

As the well-separated nanorods continue to grow beyond ~800 nm in height, they start to form new architectures. For the case of Cu, bridging occurs but nanorods generally remain separated. In contrast, nearly complete merge of nanorods occurs without the heterogeneous nucleation sites (Fig. 4a inset). For the case of Au, branching has occurred beyond ~800 nm, but the small diameter and the separation of nanorods both persist. In contrast, a dense columnar Au film grows when the substrate is a regular Si {100} substrate with native oxide (Fig. 4b inset).

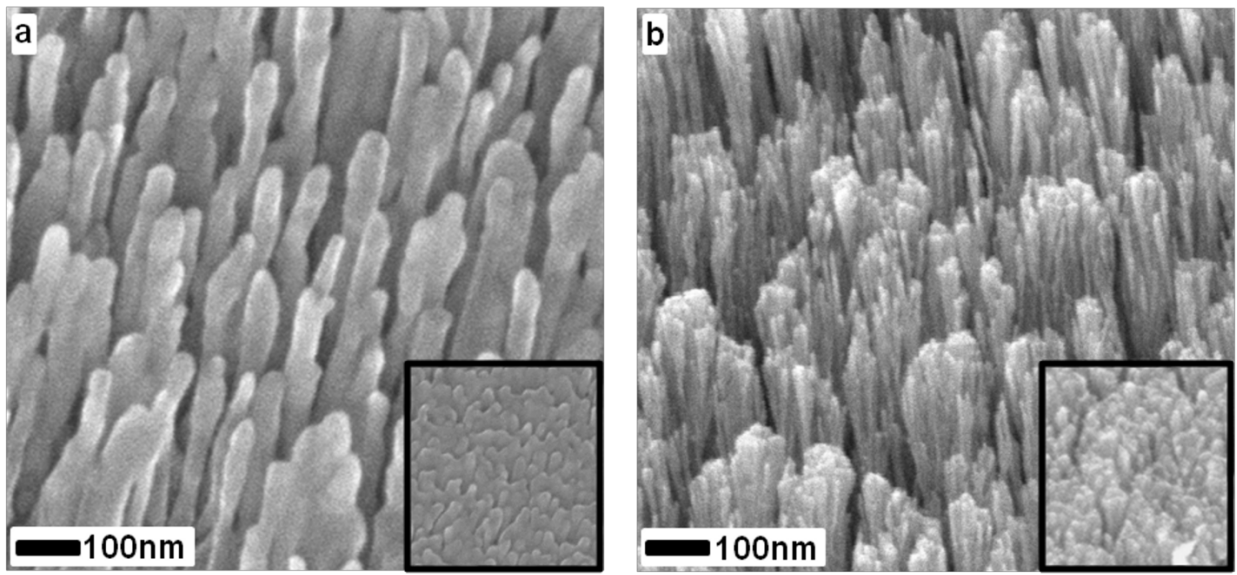

**Figure 4: Experimental results of bridged/branched nanorods.** SEM images of **a**, Cu and **b**, Au nanorods at a later stage when nanorods are about 1000 nm long; the insets with the same scale show surface morphologies of nanorods when conventional substrates are used. The incidence angle of deposition flux is 88°, the deposition rate is 0.1 nm/s, and the substrate temperature is estimated to be 250 K.

In summary, we have formulated a closed-form theory of the smallest diameter of metallic nanorods, verified the theory using LKMC simulations and validated it using previous experiments. Further, using the theory-guided PVD experiments, we have realized well-separated Cu nanorods of ~20 nm in diameter and well-separated Au nanorods of ~10 nm in diameter. These Au nanorods are the smallest well-separated metallic nanorods that have ever been reported using PVD.

**Acknowledgement:** The authors acknowledge the financial support of US DoE Office of Basic Energy Science (DE-FG02-09ER46562), and access to user facilities at the Center for Integrated NanoTechnologies at Los Alamos and Sandia National Laboratories.

**Author contributions:** X.B.N. performed the theoretical formulation and LKMC simulations, S.P.S. performed the experiments, H.C.H. designed the project, and J.K.B. & A.M. advised in the setup of experiments. All authors contributed to data analysis and editing of the manuscript.

**Additional information:** Supplementary information is available in the online version of the paper. Reprints and permissions information is available online at www.nature.com/reprints. Correspondence and requests for materials should be addressed to H.C.H.

**Competing financial interests:** The authors declare no competing financial interests.

**Figure Legends:**

**Figure 1: Model of growing the smallest nanorods. a**, Schematic of the two modes of nanorod growth, with mode II giving rise to the smallest nanorods; and **b**, evolution of a nanorod as a function of time for mode II.

**Figure 2: Theory of the smallest diameter $L_{min}$ of nanorods. a**, The theoretical distribution $S_n(L)$ for various numbers of layers $n$ in height; the inset shows a comparison of the numerical solution, the closed-form expression, and LKMC simulation results under complete geometrical shadowing as a function of $(v_{3D}/F)^{1/5}$. **b**, LKMC simulation results under incomplete geometry shadowing as a function of $(v_{3D}/F)^{1/5}$; the separation of nanorod nuclei $L_s$ is included for comparison, and the incidence angle is 85°. The inset shows nanorods from a LKMC simulation with random nucleation. **c**, LKMC simulation results under incomplete geometry shadowing as a function of incidence angle, with either the same $F_e = 1 \cdot \sin 5°$ nm/s or the same $F = 1$ nm/s; the separation of nanorod nuclei $L_s$ is included for comparison.

**Figure 3: Experimental results of the smallest well-separated nanorods.** Scanning electron microscopy (SEM) images of well-separated **a**, Cu and **b**, Au nanorods at an early stage; the insets with the same scale show the morphologies of substrates. The incidence angle of the deposition flux is 88°, the deposition rate is 0.1 nm/s, and the substrate temperature is estimated to be 250 K.

**Figure 4: Experimental results of bridged/branched nanorods.** SEM images of **a**, Cu and **b**, Au nanorods at a later stage when nanorods are about 1000 nm long; the insets with the same scale show surface morphologies of nanorods when conventional substrates are used. The incidence angle of deposition flux is 88°, the deposition rate is 0.1 nm/s, and the substrate temperature is estimated to be 250 K.